\documentstyle[12pt,titlepage]{article}
\newcommand{\ba}{\begin{array}}
\newcommand{\ea}{\end{array}}
\newcommand{\bd}{\begin{displaymath}}
\newcommand{\ed}{\end{displaymath}}
\newcommand{\be}{\begin{equation}}
\newcommand{\ee}{\end{equation}}
\newcommand{\bea}{\begin{eqnarray}}
\newcommand{\eea}{\end{eqnarray}}

\def\bra{\langle}
\def\ket{\rangle}

\def\a{\alpha}
\def\b{\beta}
\def\g{\gamma}

\def\e{\epsilon}

\def\m{\mu}
\def\n{\nu}
\def\G{\Gamma}

\def\p{\pi}

\begin{document}
\begin{titlepage}
\vspace*{0.5truein}

\begin{center}
{\Large\bf CAN WE IDENTIFY A LIGHT NEUTRALINO IN B-FACTORIES ?  \\[0.5truein]}
{\large Rathin Adhikari$^1$ and Biswarup Mukhopadhyaya$^2$\\}
Mehta Research Institute \\
10 Kasturba Gandhi Marg\\
Allahabad - 211 002, INDIA \\
\end{center}
\vskip .5cm

\begin{center}
{\bf ABSTRACT}\\
\end{center}
If a light gaugino sector exists, then the lightest supersymmetric particle
(LSP) has a chance of being pair-produced in rare B-decays. As a
consequence of neutral flavour violation in most supersymmetric models,
such decays can occur at the tree-level and reinforce the channels
$B \longrightarrow K(K^{*}) + invisible$. We discuss how a study of
such decay spectra in B-factories can help us either identify or exclude
a light LSP.
\hspace*{\fill}
\vskip .5in

\noindent
$^{1}$E-mail : ~~rathin@mri.ernet.in  \\
$^{2}$E-mail :biswarup@mri.ernet.in

\end{titlepage}

\textheight=8.9in

   Since supersymmetry (SUSY) \cite{susy} is one of the most appealing
prospects in our quest for physics beyond the standard model, all possible
ways to
uncover (or rule out) its existence are worthy of attention, even if it be
in relatively inprobable corners. Though the lower bounds on squark and
gluino masses as announced by the CDF collaboration are 126
GeV and 141 GeV respectively \cite{cdf}, they are based on certain
assumptions about superparticle decays and SUSY parameters. Relaxation of
such limits cannot thus be altogether excluded \cite{pdg}. In particular,
there is a persistent, although controversial, claim that a window in the
parameter space with a light gluino is still open \cite {lg1,lg2}. If that
indeed be the case , then a gluino in the mass range of, say, 2.5 - 5 GeV
will cause a squark to decay directly into it. The gluino will
subsequently decay into the lightest supersymmetric particle (LSP),
supposedly even lighter in an R-parity conserving SUSY. As a result the
missing transverse momenta $(\not{P_T})$ associated with the LSP  are
considerably degraded and cannot survive the $\not{P_T}$-cuts employed in
conventional SUSY searches. Therefore , this scenario also relaxes the
squark mass limits \cite{lg1}. Further
motivations for a light gluino have come from
the observation \cite{clav} that the values of the strong coupling $\alpha_s$
at low
and high energies are in better agreement with theory if a low-mass,
electrically neutral coloured fermion is present. Attempts have been made to
demonstrate the plausibility of such a scenario by proposing SUSY models where
the gauginos acquire their masses radiatively \cite{mas}. On the
phenomenological side,
recent studies include the implications of a light, long-lived gluino in
strong processes \cite{far}, constraints on light gluinos from
electroweak precision
tests  \cite{ag1} and those coming from radiative b-decay \cite{ag2}.

Here we focus on such a scenario from another angle; namely, we try to
extract signatures {\it of the light LSP} from the kinematics of rare B-meson
decays \cite{rareb}. We assume the LSP to be the lightest neutralino
($\chi^0_1$) in this
case. Then the flavour-changing neutral current (FCNC) decays
$B\longrightarrow K(K^{*}) \chi^0_1 \chi^0_1$ will give the same
observable final states as $B\longrightarrow K(K^{*}) \nu \overline{\nu}$.
However, the decay spectra for the SUSY and standard model (SM) processes are
going to have different shapes. The net observed variation of
$d\Gamma/dE_{K(K^{*})}$
with the K(K$^*$)-energy will be a result of superposition of the two types of
final states, leading to a distribution with a kink.
The position of the kink and
the distortion to the spectrum relative to the purely SM case depends on
the mass of the LSP. Some earlier studies had suggested the application of
a similar effect in Kaon decay to unmask a massive tau-neutrino or a
photino \cite{deshpb}.
We have also recently discussed the implications of such spectral
distortion in the context of decays of a heavy Higgs boson \cite{br}.

There are several advantages in looking for a light LSP in B-decays. First,
as we shall see, an LSP in the mass range $0.5 MeV-1 GeV$ produces a kink in
a conspicuous part of the spectrum, thereby making the distortion rather
obvious. If we explore the window in $m_{\tilde {g}} = 3-5 GeV$, then the
LSP is likely to lie in the above range. Secondly, B-decays are dominated
by short-distance physics, so that quark-level estimates are reliable.
And finally, with a number of B-factories being designed for the near future,
the prospect of producing $10^{7-8}$ $B\overline{B}$-pairs are
realistic \cite{bfac}. Thus
one may aspire to have a sufficient number of B's at the threshold so
that their rare decays can be under scrutiny in the rest frame.

Both the decays $B\longrightarrow K \chi^0_1 \chi^0_1$ and
$B\longrightarrow K^{*} \chi^0_1 \chi^0_1$ are driven by the quark-level
process $b\longrightarrow s \chi^0_1 \chi^0_1$. In the supersymmetric
standard model, there is the interesting
posibility of tree-level flavour violation in quark-squark-neutralino
(or quark-squark-gluino) interactions \cite{fcnc}. This can occur because
the quark
and squark mass matrices are not simultaneously diagonal. In a basis where
the $3 \times 3$ down-quark mass matrix is diagonal, the $6 \times 6$
down-squark mass matrix is given by

\bea
M_{\tilde{d}}^2 =
\left( \ba{lr}
m_L^2 \, {\bf{1}} + m_{\hat{d}}^2+c_0 \, K \, m_{\hat{u}}^2 \, K^{\dag}    &
\!\!{\large{0}}  \\
\:\:\:{\large{0}}  & m_R^2 \, {\bf{1}} +m_{\hat{d}}^2
\ea \right)
\eea

\noindent
where $m_{\hat{d}}, m_{\hat{u}}$ are the diagonal down-and up-quark mass
matrix respectively, and K is the Kobayashi-Maskawa matrix. $m_L$ and
$m_R$ are the flavour-blind SUSY breaking parameters that set the scale
of squark masses. We can put $m_L = m_R$ without any loss of generality.
The term proportional to  $m_{\hat{u}}^2$ occurs due to radiative corrections
induced by Yukawa coupling with charged Higgsinos. Such a term is particularly
important in a model based on N=1 supergravity (SUGRA), as the mass parameters
evolve from the high SUGRA breaking scale to the scale of the residual
global SUSY breaking at a lower energy.
Because of this term,  $m_{\tilde{d}}^2$
cannot be simultaneously diagonal with  $m_{\hat{d}}^2$. As a result,
quark-squark-neutralino (or gluino) interactions
can violate flavour. The flavour
mixing in the down sector is controlled essentially by the top-quark mass,
the Kobayashi-Maskawa matrix and the parameter $c_0$. The calculation of
$c_0$ is model-dependent; we shall treat it here as a phenomenologcal
input that needs to be restricted by SUSY contributions to various FCNC
processes.

In the above framework (where we have also neglected the mixing between
left-and right-squarks), the quark-squark-neutralino
$(q-{\tilde{q}}-\chi^{0}_{1}, i=1-4)$ coupling in the down sector is

\bea
{\cal L}_{q\tilde{q}\chi_i^0}  =
-{\sqrt{2}} \, g \, \sum_{ij} \, {\tilde q}_j \, {\bar{\chi}_i^0} \, \left[
\, \tan{\theta_w} \, e_j N_{i2}^\star \,
\Gamma_{jk} \,  {{1-\g_5} \over 2}  +{\delta}_{jk} \, \left( T_{3j}
\, N_{i2} - \tan{\theta_w} \left( T_{3j} -e_j
\right) \, N_{i1} \right) \,\right. \nonumber \\
\left. {{1+ \g_5} \over 2} \right] q_k \, + \, h.c.
\eea

\noindent
where $\Gamma_{jk}$ is the (jk)-th element of the unitary matrix that
diagonalises the upper $3 \times 3$ block of $m_{\tilde{d}}^2$ in equation(1).
N is the neutralino mixing matrix, and $T_{3j}$ the third component of the
the isospin of the j-th flavour.
In the absence of left-right mixing, only the first term in (2) is
relevant for flavour-violating interactions.

For our calculations, the element $\Gamma_{23}$ will be important. We have
used $m_t =170 GeV$. For such a top-quark mass, the third term in the
upper-left block of $m_{\tilde{d}}^2$ is important from the viewpoint
of diagonalisation, so that for a not-too-small value of $c_0$, the
elements of $\Gamma$ are close to those of K in magnitude. We have
parametized $\Gamma_{23}$ by writing $\Gamma_{23} = c K_{23}$.  Various values
of $c_0$ and and the corresponding values of c are given in table 1.

The two tree-level graphs (plus those with the momenta of the neutralinos
interchanged) shown in figure 1 contribute to
$b\longrightarrow s \chi^0_1 \chi^0_1$. For numerical calculations, we
confine ourselves to a situation where the gluino mass is in the range
3-5 GeV. In such a case, the other parameters in the SUSY sector have to be
compatible with the LEP-I results \cite{lep1}. Under such
circumstances, it is straightforward
to verify that the LSP is dominated overwhelmingly by the photino. This is
easily demonstrated if, for example, we adhere to a scenario inspired by
Grand Unified Theories (GUT) \cite{susygut}. In such cases the only
independent inputs
apart from the gluino mass are $\m$, the Higgsino mass parameter,
and  $tan \beta$, the ratio of the two scalar vacuum expectation values.
In the above situation one gets confined to $50 GeV \leq \m \leq 100 GeV$ and
$1.0 \leq tan \beta \leq 1.5$.  Diagonalisation of the neutralino mass
matrix with such inputs reveals the complete dominance of the photino
state in the LSP. Therefore,
the interaction (2) can be used in the $\chi_{1}^0 = \tilde{\g}$
limit for our purpose [1(b)]. The amplitude for
$b(p_0)\longrightarrow s(p_3) \chi^0_{1} (p_2) \chi^0_{1} (p_1)$ can be
expressed as

\bea
{\cal M} = {e^2 \; c \; V_{23} \over 9}\;  \left [  {1 \over  [{(p_0
-p_1)}^2 \;
-{m_{\tilde{q}}}^2]}
\; {\bar u}(p_1)\; (1-\g_5)\;b(p_0)\; {\bar
s}(p_3)\;(1+\g_5)\; v(p_2) \; - \right. \nonumber \\  \left. {1 \over
[{(p_0 -p_2)}^2 \;
-{m_{\tilde{q}}}^2]}
\;{\bar u}(p_2)\;
(1-\g_5)\;b(p_0)\; {\bar
s}(p_3)\;(1+\g_5)\; v(p_1)  \; \right ]
\eea

\noindent
where $m_{\tilde{q}} $ is the assumed common mass of the ${\tilde s}$ and
$ {\tilde b}$ squarks.
In order to factor out the hadronic part of the amplitude, one needs to do
a Fierz transformation on (3), which gives

\bea
{\cal M} = {e^2 \; c \; V_{23} \over 18}\;  \left [  {1 \over  [{(p_0
-p_1)}^2 \;
-{m_{\tilde{q}}}^2]}
\; {\bar
s}(p_3)\; \g^\m \;(1-\g_5)\;b(p_0)\;{\bar u}(p_1)
\;\g_\m \;(1+\g_5)\; v(p_2) \; -
\right. \nonumber \\  \left. {1 \over
[{(p_0 -p_2)}^2 \;
-{m_{\tilde{q}}}^2]}
\;{\bar
s}(p_3)\;\g^\m \;
(1-\g_5)\;b(p_0)\; {\bar u}(p_2)\;\g_\m \;(1+\g_5)\; v(p_1)
\; \right ]
\eea

To proceed further, one has to obtain the hadronic matrix elements for the
above quark current. Using a common parametrization for the matrix elements
for rare B-decays \cite{deshb}, one can write

\be
\bra K(p_3) |\; \bar{s} \g_\m  b|\;B(p_0) \ket =
f_+(q^2) {(p_3+p_0)}_\m + f_-(q^2) q_\m \quad ,
\ee
\be
\bra K(p_3) |\; \bar{s}\;  \g_\m \g_5 \; b|\;B(p_0) \ket =\; 0
\ee
\bea
\bra K^\star (p_3) |\; \bar{s}\;  \g_\m \;(1- \; \g_5\;) \; b|\;B(p_0)
\ket =\; i \; \e_{\m \n \a \b } \; \e^\n (p_3) \; {(p_0 +p_3)}^\a \; q^\b
V(q^2) \; - \nonumber \\  \e_\m (p_3)\; \left[ \; m_B^2 -\;m_{K^\star}^2 \;
\right] \; A_0 (q^2)\; - (\e \; . \; q) \; {(p_0 +p_3)}_\m \; A_+ (q^2) \;
- \; (\e \; . \; q )\; q_\m \; A_-(q^2)
\eea

\noindent
where $q_\m \; = \; {(p_0 -p_3)}_\m$ and
$\e_{\m}(p_3)$ is the polarization vector for the $K^*$.
Our results are based upon numerical values of the various form-factors
(and pole fits for their momentum-transfer dependence) obtained from the
relativistic quark model of reference \cite{wsb}. These form-factors have been
computed
in the literature using other models, too \cite{bmod};
the question of matching such
calculations with the general relationship predicted by heavy quark effective
theories have also been discussed \cite{hqet}. However, the
uncertainties due to
model-dependence do not affcet the general features of our results. It should
also be noted that we have not taken QCD corrections into account here. Though
such corrections moderately alter the decay rates \cite{qcd}, the key
featurs of the
spectral distortions should not be affected, since at the lowest
order electroweak level, the SUSY and standard model effective interactions
have the same operator structure.

The differential decay rates of our interest are given by
\bea
{d  \G \over d E_{K\;(K^\star)}} \; = \; {1 \over 64 {     \p}^3 \;
m_{B}} \;\int_{E_1(min)}^{E_1(max)} \;\;{|{\cal M}|}^2 \; dE_1
\eea

where ${|{\cal M}|}^2$      is the squared matrix element and
\bea
E_1(max) = {1 \over 2}\; \left[ (m_{B} - E_{K\;(K^\star)}\; + {\bf p_3}\;
{\sqrt{ 1-\;4
m_{\chi^0_1}^2 / q^2}}\;\; \right]
\eea

\bea
E_1(min) = {1 \over 2}\; \left[ (m_{B} - E_{K\;(K^\star)}\; - {\bf p_3}\;
{\sqrt{ 1-\;4
m_{\chi^0_1}^2 / q^2}}\;\; \right]
\eea

$
\; \; \; \; \; \; \; \; q^{2}=\;m_{K\;(K^\star)}^2+\;m_{B}^2-2 \;m_{B}\;
E_{K\;(K^\star)}
$
\hfill (11a)

\noindent
$
\; \; \; \; \; \; \; \; \; \; \; \;
\; \; {\bf p}_{3}^2=\;E_{K\;(K^\star)}^2-\;m_{K\;(K^\star)}^2
$
\hfill (11b)

\hspace*{\fill}

\noindent
$E_{K\;(K^\star)}$ is the ${K\;(K^\star)}$ energy in the rest
frame of decaying B  and  its kinematically allowed range is

\noindent
$ \; \; \; \; \; \; \; \; \; \; \; \; \; \; m_{K\;(K^\star)} \leq
E_{K\;(K^\star)}
\leq (m_B^2
+m_{K\;(K^\star)}^2 -
4 m_{\chi^0_1}^2)/2 m_B
$
\hfill (11c)

\hspace*{\fill}

\noindent
Further, one has to add
the rates for the
SUSY process with that for
$\Sigma B\longrightarrow K(K^{*}) \nu_i \overline{\nu_i}$ which occurs via
triangle as well as box diagrams \cite{soni}.

The numerical results are shown in figures 2-5. For the $K^*$ final states,
the polarizations have been summed over. The fact that the standard
model final states consist of $\nu \overline{\nu}$ pairs explains why the
distributions are not vitiated by peaks due to the $J/\psi$ and $\psi{\prime}$
resonances. We have drawn the graphs for $m_{\tilde{q}} = 100 GeV$ which is
easily allowed in this scenario. Two sets of graphs each for the $K$ and
$K^*$ final states are presented in order to demostrate the dependence of
the rates on the parameter $c_0$. Evidently, with even quite  conservative
choices for $c_0$ one can notice distortions to
the spectrum over a considerable region of the parameter space. The effect
becomes less and less obvious with increasing squark mass, and is
barely perceptible for $B\longrightarrow K \chi^0_1 \chi^0_1$ with
$c_{0} \approx .01$, $m_{\tilde{q}} = 500 GeV$. Also,
the response to a variation in the mass of the LSP
in the region  $0.4-1 GeV$ is manifest. A few hundred events in a B-factory
should suffice to explore this kind of a distortion.

It is to be noted that while the differential decay rate for
$\Sigma(B\longrightarrow K \nu_{i} \overline{\nu_{i}})$ increasess
monotonically  with $E_K$,
it dips after an initial rise in the case of
$\Sigma(B\longrightarrow K^{*} \nu_{i} \overline{\nu_{i}})$.
This is because the transverse
component of $K^*$ has an important role in the rate. For it,
the upper end of the phase space (corresponding to the maximum
allowed value of $E_{K^*}$) corresponds to a configuration that is
disfavoured by helicity conservation. The corresponding distortions
to the spectrum caused by the LSP are less conspicuous than in the case of
decays into a K,
although the overall rate for the former is higher by about an order of
magnitude.

The graphs also clearly portray the fact that with light LSP's the total rates
for both the decays can be jacked up by as much as an order of magnitude due
to presence of a light LSP. At least over a certain amount of parameter space
such an overall rate enhancement can be taken as a positive signal of
such an LSP.

To conclude, if $10^{7-8}$ $B \overline{B}$-pairs are produced in a B-factory
per year, then the tree-level flavour violating interaction in SUSY models
strongly affect the decay patterns in $B \longrightarrow K + nothing$ and
$B \longrightarrow K^{*}+ nothing$, provided that a light LSP is present.
This can be, in a somewhat model-dependent way, a pointer to the light gluino
as well. Therefore, a detailed investigation of the above types
of decays in B-factory experiments are going to help one in constraining the
light sparticle scenario to a large extent.

\newpage
\centerline{\large {\bf Table caption}}
\noindent
Table 1: Different values of the parameter $c_0$ and the
corresponding values of c, for $m_{\tilde{q}} = 100\;GeV$.
\newpage
\centerline {\large {\bf Figure Captions}}

\hspace*{\fill}

\hspace*{\fill}

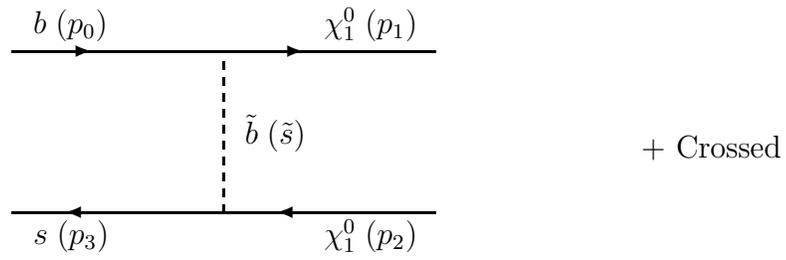
Figure 1:

\noindent
The tree-level contributions to $b\longrightarrow s \chi^0_1 \chi^0_1$.
In addition there will be crossed diagrams where the four-momenta of
the LSP's are interchanged.

\vskip .25in

Figure 2:

\noindent
The differential decay rates for $B {\longrightarrow} K + nothing$
for $m_{\tilde{q}}\; = 100 GeV, \; c = 0.1$. The solid, dotted and short-dashed
curves correspond to three LSP masses expressed in GeV. The long-dashed
curve below is for the purely standard model case with three massless
neutrinos.

\vskip .25in

Figure 3:

\noindent
Same as figure 2, but with c = 0.5.

\vskip .25in

Figure 4:

\noindent
The differential decay rates for $B {\longrightarrow} K^{*} + nothing$,
with the same choice of parameters as in figure 2.

\vskip .25in

Figure 5:

\noindent
Same as in figure 4, but with c=0.5.

\newpage

\hspace*{\fill}

\hspace*{\fill}

\hspace*{\fill}

\hspace*{\fill}


\begin{table}[h]
\begin{center}
\begin{tabular}{||l|l|l|l||l}
\hline
 $c_0$           &  0.01 & $0.001$ & $0.0001$
\\ \hline
 $c$     &  0.9& 0.5 & 0.1
\\ \hline
\end{tabular}

\hspace*{\fill}

\hspace*{\fill}

\hspace*{\fill}

{\Large Table 1}
\end{center}
\end{table}

\hspace*{\fill}

\hspace*{\fill}

\newpage

\hspace*{\fill}

\hspace*{\fill}

\hspace*{\fill}

\hspace*{\fill}


\begin{picture}(150,250)(-100,-160)
\thicklines
\put(22,4){\line(1,0){20}}
\put(52,4){\vector(-1,0){10}}
\put(52,4){\line(1,0){70}}
\put(132,4){\vector(-1,0){10}}
\put(132,4){\line(1,0){50}}
\put(22,65){\vector(1,0){30}}
\put(52,65){\line(1,0){50}}
\put(102,65){\vector(1,0){30}}
\put(132,65){\line(1,0){50}}
\put(102,4){\line(0,1){3}}
\put(102,10){\line(0,1){3}}
\put(102,16){\line(0,1){3}}
\put(102,22){\line(0,1){3}}
\put(102,28){\line(0,1){3}}
\put(102,34){\line(0,1){3}}
\put(102,40){\line(0,1){3}}
\put(102,46){\line(0,1){3}}
\put(102,52){\line(0,1){3}}
\put(102,58){\line(0,1){3}}
\put(140,-8){\mbox {$\chi^0_1\;(p_2)$}}
\put(140,72){\mbox {$\chi^0_1\;(p_1)$}}
\put(30,-8){\mbox {$s\;(p_3)$}}
\put(30,72){\mbox {$b\;(p_0)$}}
\put(110,30){\mbox {${\tilde b \; ({\tilde s})} $}}
\put(260,25){\mbox {+ Crossed}}
\put(100,-70){\mbox {\Large FIG. 1}}
\end{picture}


\newpage


\begin{thebibliography}{23}

\bibitem{susy} For reviews see, for example, (a) H. P. Nilles, Phys. Rep.
{\bf 110} (1984) 1; (b) H. Haber and G. Kane, Phys. Rep. {\bf 117} (1985) 75.

\bibitem{cdf} F. Abe {\it et al.}, Phys. Rev. Lett. {\bf 69} (1992) 3439.

\bibitem{pdg} Review of Particle Properties, Phys. Rev. {\bf D50} (1994) S1,
and references therein.

\bibitem{lg1} C. Albajar {\it et al.}, Phys. Lett. {\bf B198} (1987) 261.

\bibitem{lg2} P. Tuts {\it et al.}, Phys. Lett. {\bf B186} (1987) 233; J. Lopez
{\it et al.}, Phys. Lett. {\bf B313} (1993) 241; C. Carlson and M. Sher,
Phys. Rev. Lett. {\bf 72} (1994) 2686.

\bibitem{clav} L. Clavelli, Phys. Rev. {\bf D46} (1992) 2112; L. Blumlein
and J. Botts, Phys. Lett. {\bf B325} (1994) 190.

\bibitem{mas} G. Farrar and A. Masiero, Rutgers University Report RU-94-38
(1994).

\bibitem{far} G. Farrar, Rutgers University Report RU-94-35 (1994).

\bibitem{ag1} G. Bhattacharyya and A. Raychaudhuri, Phys. Rev. {\bf D49}
(1994) R1156.

\bibitem{ag2} G. Bhattacharyya and A. Raychaudhuri, CERN Report
CERN-TH-7245/94 (1994).

\bibitem{rareb} For a status update, see, for example, D. Kim, Cornell
University Report CLNS 93/1254 (1993); N. Deshpande, Invited Talk presented
at the XXVII International Conference on High Energy Physics, Glasgow (1994).

\bibitem{deshpb} N. Deshpande and G. Eilam, Phys. Rev. Lett. {\bf 53} (1984)
2289; J. Nieves and P. B. Pal, Phys. Rev. {\bf D32} (1985) 1849.

\bibitem{br} R. Adhikari and B. Mukhopadhyaya, ICTP Report IC/94/174 (1994).

\bibitem{bfac} See, for example, S. Stone (ed), {\it B-decays}, World
Scientific (1992).

\bibitem{fcnc} J. Donoghue {\it et al.}, Phys. Lett. {\bf B128} (1983) 55;
M. Duncan, Nucl. Phys. {\bf B221} (1993) 285; A. Bouquet {\it et al.}, Phys.
Lett. {\bf B148} (1984) 69.

\bibitem{lep1} A. Datta {\it et al.}, Z. Phys. {\bf C54} (1992) 513;
H. Baer {\it et al.}, Phys. Rev. {\bf D47} (1993) 1062.

\bibitem{susygut} G. Ross and R. Roberts, Nucl. Physics {\bf B377} (1992) 571;
L. Ibanez and G. Ross in {\it Perspectives on Higgs Physics}, G. Kane (ed),
World Scientific (1993).

\bibitem{deshb} N. Deshpande and J. Trampetic, Phys. Rev. Lett. {\bf 60}
(1988) 2583.

\bibitem{wsb} M. Baur {\it et al.}, Z. Phys. {\bf C29} (1985) 637, {\it ibid},
{\bf C34} (1987) 103.


\bibitem{bmod} N. Isgur {\it et al.}, Phys. Rev. {\bf D39} (1989) 799; N. Isgur
and M. Wise, Phys. Lett. {\bf B237} (1990) 527; W. Jaus and D. Wyler, Phys.
Rev. {\bf D41} (1989) 3405; G. Greub {\it et al.}, University of Zurich Report
ZU-TH-25/94 (1994).

\bibitem{hqet} H. Georgi, Phys. Lett. {\bf B240} (1990) 447,
{\it ibid}, {\bf B247} (1990) 399; N. Isgur and M. Wise,
Phys. Rev. {\bf D42} (1990) 2388;  M. Neubert, SLAC Report SLAC-PUB-6263
(1993);
F. Close and A. Wambauch, Nucl. Phys. {\bf B412} (1994) 169; M. Voloshin,
University of Minnesota Report TPI-MINN-94/18-T (1994).

\bibitem{qcd} M. Grinstein {\it et al.}, Nucl. Phys. {\bf B319} (1989) 271.

\bibitem{soni} W. Hou {\it et al.}, Phys. Rev. Lett. {\bf 58} (1987) 1608.



\end{thebibliography}
\end{document}